\documentclass{optica-article}

\journal{opticajournal} 

\articletype{Research Article}
\usepackage{xcolor}
\usepackage{lineno}
\usepackage{hyperref}

\begin{document}

\title{Linewidth narrowing and wideband frequency modulation of a DBR laser}

\author{Jack Roth,\authormark{1,*} Andrew Christensen,\authormark{1} Madeline Bernstein,\authormark{1} Yuno Iwasaki,\authormark{1} Hana Lampson,\authormark{2} and Holger Mueller\authormark{1}}

\address{\authormark{1}Department of Physics, University of California, Berkeley, CA 94720, USA}
\address{\authormark{2}Department of Physics, Harvard University, Cambridge, MA 02138, USA}
\email{\authormark{*}jack\_roth@berkeley.edu} 


\begin{abstract*}
We present a scheme to phase-lock a $240\,\mathrm{mW}$, $852\,\mathrm{nm}$ distributed Bragg reflector (DBR) laser to a fixed-frequency narrow-linewidth laser with a rapidly tunable frequency offset near $9\,\mathrm{GHz}$. The phase-lock is accomplished by electronic feedback on the beatnote between the two lasers. The frequency offset can be swept $200\,\mathrm{MHz}$ in $300\,\mathrm{\mu s}$, limited by the feedback loop bandwidth, allowing for its use in complex cooling and state preparation schemes needed in atomic physics experiments. Additionally, we find that the phase-lock reduces the linewidth of the DBR laser below its natural linewidth of $\sim400\,\mathrm{kHz}$ to $\sim100\,\mathrm{kHz}$.
\end{abstract*}


\section{Introduction}

Atomic, molecular, and optical physics experiments often require swept laser frequencies to, e.g., generate Bloch oscillations \cite{Clad2009,Ferrari2006,Peik1997}, moving optical molasses \cite{Treutlein2001}, or switch from magneto-optical trapping to sub-Doppler optical molasses cooling \cite{Lett1988}. These sweeps may require the precision of a phase-lock, hundreds of MHz range, and millisecond speeds. Compared to optical modulators, directly sweeping a frequency-stabilized (“locked”) laser by changing its lock point reduces optical complexity and keeps the full output power of the laser available. Distributed Bragg reflector or distributed feedback diode lasers (DBRs and DFBs) offer a large mode-hop free tuning range and robustness compared to external cavity diode lasers (ECDLs), but their larger linewidth complicates frequency stabilization, especially phase-locking.
Using photonic integrated circuits with a large feedback bandwidth \cite{Ishida1991} has achieved a phase-lock between two custom fabricated DBRs \cite{Ristic2010}. A DFB was phase-locked \cite{Friederich2010} without a photonic integrated circuit, although the frequency modulation capabilities were not explored. Other authors have narrowed the linewidth of DBRs and DFBs using electronic feedback based on a reference cavity \cite{reduced_linewidth_cavity,Ohtsu_1985} or an interferometer with unbalanced path lengths \cite{reduced_linewidth_interferometer_1,reduced_linewidth_interferometer_2,reduced_linewidth_interferometer_3,reduced_linewidth_interferometer_4}. However, achieving a frequency-agile phase-lock of a DBR laser with commonly available components remains challenging.

Here, we phase-lock a commercial DBR with linewidth narrowing and wideband frequency modulation by using only commercially available components, taking advantage of a DBR with $<1\,\mathrm{MHz}$ linewidth. The DBR frequency can be swept across $200\,\mathrm{MHz}$ within $300\,\mathrm{\mu s}$, sufficient to address different excited hyperfine states in cesium. We offset the DBR from the reference by about $9\,\mathrm{GHz}$ and sweep it multiple times within a $20\,\mathrm{ms}$ period to generate a magneto-optical trap followed by optical molasses and Raman sideband cooling. The phase-lock narrows the DBR linewidth from $\sim400\,\mathrm{kHz}$ to $\sim100\,\mathrm{kHz}$, about twice the linewidth of the reference. This describes a simple and versatile tool for generating and modulating laser frequencies in atomic physics labs. 

\section{Experimental setup}

\begin{figure}[t]
\centering\includegraphics[width=\textwidth]{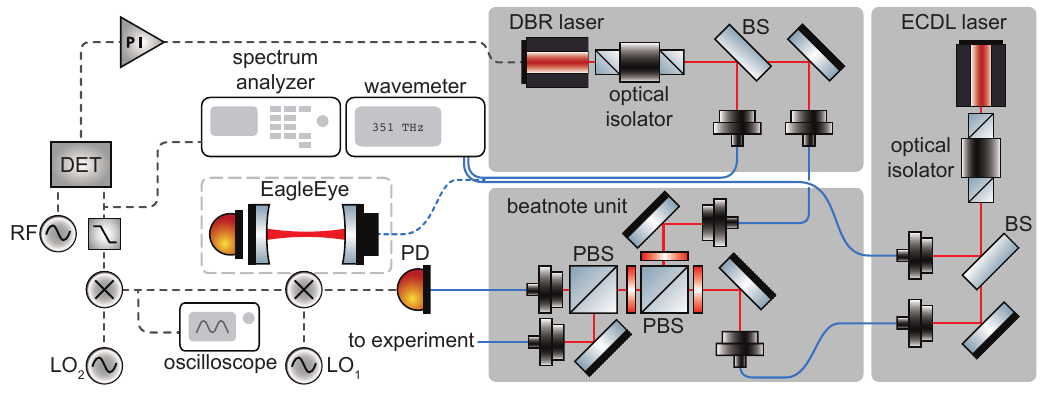}
\caption{The ECDL reference laser and the DBR lasers are fiber coupled and sent to the beatnote unit. The beatnote is recorded on a fast photodiode and then downconverted, first by $f_{\text{LO}_1}$ and then again by $f_{\text{LO}_2}$. The downconverted beatnote is low-pass filtered and then sent to the phase-frequency detector (labeled DET, detailed in Fig.~(\ref{fig:phase_freq_schematic})). $f_\text{RF}$ is used as the phase-frequency detector reference frequency. The error signal generated by the phase detector circuit is sent to a PI controller feeding back on the DBR laser current (labeled PI, detailed in Fig.~(\ref{fig:feedback_schematic})). This schematic omits additional optics present on the output of the ECDL reference laser used to stabilize it to a cesium vapor cell.}
\label{fig:optics_schematic}
\end{figure}

The setup is pictured in Fig.~(\ref{fig:optics_schematic}). The reference for our phase-locked DBR is a $40\,\mathrm{mW}$ ECDL (AOSense AOS-IF-ECDL-852) stabilized to a cesium vapor cell. The light is split into three fibers. Approximately $3\,\mathrm{mW}$ are sent to a cesium vapor cell to stabilize the ECDL  to the $F=3\to F'=4$ transition in the cesium $D_2$ line via frequency modulation spectroscopy (not pictured). Another $3\,\mathrm{mW}$ is sent to a fiber used for frequency diagnostics; the remaining $34\,\mathrm{mW}$ is sent to the beatnote unit. The DBR laser (Photodigm 852.347DBRH-L-TO8) outputs up to $240\,\mathrm{mW}$, and is nominally $5.4\,\mathrm{MHz}$ red-detuned from the $F=4\to F'=5$ transition in the cesium $D_2$ line. A few milliwatts are picked off and sent into a fiber used for frequency diagnostics. The remaining power is sent to the beatnote unit.

The beatnote unit overlaps the modes of the ECDL and DBR laser beams. First, beams from both lasers are combined on a polarizing beamsplitter cube (PBS) with their polarizations set so that both beams leave via the same output port. A second PBS projects both beams onto the same polarization. Both outputs of the second PBS are coupled into separate fibers. The relative power from the ECDL and DBR lasers is controlled with a waveplate between the two PBSs. One fiber is sent to a fiberized $15\,\mathrm{GHz}$ bandwidth photodiode (Optilab SPD-15-A-DC) used to measure the beatnote. The other fiber, which now contains both frequencies required to generate a magneto-optical trap, can be sent directly to the experiment or can be further amplified if more power is required.

\begin{figure}[p]
\centering\includegraphics[width=\textwidth]{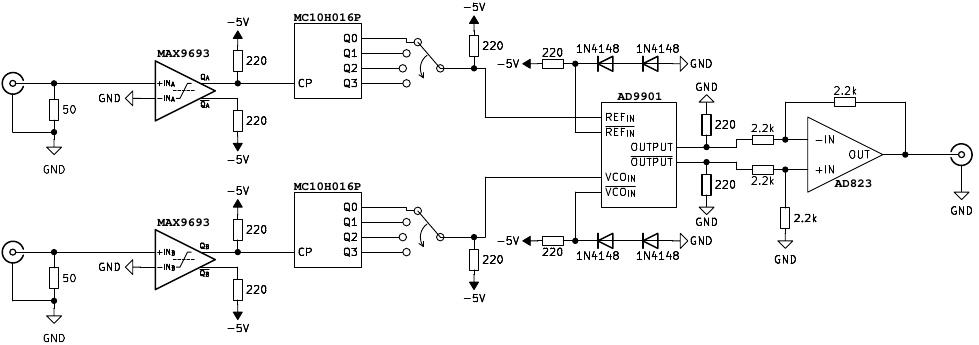}
\caption{The two input ports of the phase-frequency detector are terminated with $50\,\Omega$ and are each fed into a MAX9693 ECL comparator, which digitizes the signal. The output of each comparator is fed into a MC10H016P binary counter, which acts as a frequency divider. By selecting the desired output of the MC10H016P the frequency can be divided by 2, 4, 8, or 16. The results in this paper are performed with the division factor set to 2 on all counters (resulting in a division factor of 4 on each input signal). The divided signals are then sent to an AD9901 phase-frequency detector, which outputs a voltage proportional to the phase difference between its two inputs. If the two inputs have different frequencies the output will rail either high or low, depending on which input frequency is higher. To convert the differential ECL signal into a single ended signal an AD823 operational amplifier configured as a differential amplifier is used. Two diodes and a $220\,\Omega$ resistor are used to fix the AD9901 ECL port voltages at approximately $-2\,\mathrm{V}$.}
\label{fig:phase_freq_schematic}
\end{figure}

The phase-frequency detector outputs a voltage proportional to the phase difference between its two inputs, the doubly downconverted beatnote $f_\text{down}$ and a fixed reference signal $f_\text{RF}=150\,\mathrm{MHz}$. The output is used as the PI controller error signal. The phase-lock therefore seeks to stabilize the beatnote frequency to:
$$f_\text{PD}=\mp f_\text{RF}+f_{\text{LO}_1}+f_{\text{LO}_2},$$
where the sign of $f_\text{RF}$ is determined by the feedback polarity. In the scheme presented here the sign is positive. A detailed schematic of the phase-frequency detector is provided in Fig.~(\ref{fig:phase_freq_schematic}).

\begin{figure}[p]
\centering\includegraphics[width=\textwidth]{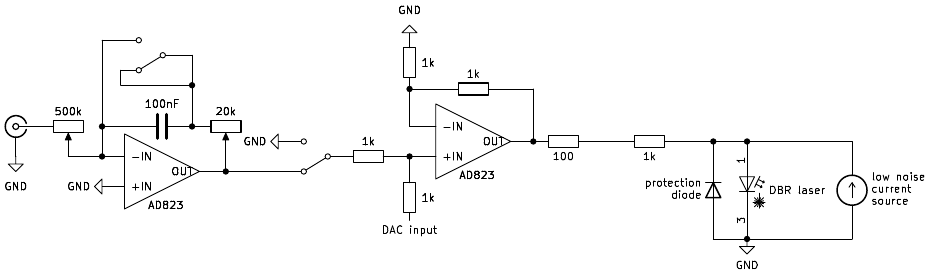}
\caption{The PI controller stage consists of a single inverting AD823 operational amplifier with a resistor and capacitor in series along the returning path. $500\,\mathrm{k}\Omega$ and $20\,\mathrm{k}\Omega$ potentiometers control the PI and P gains, respectively. A digitally controlled switch parallel to the capacitor can be used to disable the I gain leaving only the P gain. Following the PI controller stage is a second digitally controlled switch used to enable/disable the PI controller from entering the subsequent summing amplifier. The summing amplifier sums the control signal with a signal from a DAC. The DAC is scanned by a computer to bring the laser frequency close to the lock point before the PI controller is engaged. The output of the summing amplifier is injected into the laser current by bridging the connection with a $1.1\,\mathrm{k}\Omega$ resistor.}
\label{fig:feedback_schematic}
\end{figure}

The error signal from the phase-frequency detector is then fed into an operational amplifier serving as a combined proportional-integral (PI) controller, with gains that are adjustable by potentiometers. An analog switch allows the feedback to be disabled, and a summing amplifier allows the output to be swept by a digital-to-analog converter (DAC); this allows for automatic relocking. See Fig.(\ref{fig:feedback_schematic}) for a schematic of the PI controller. The output of the PI controller is then used to modulate the current of the DBR laser. The current driver for the DBR is a low-noise current source based on the design described in Ref.~\cite{libbrecht_hall_steck_driver}. In principle, one could use any current source with sufficiently low noise that the laser linewidth is not broadened beyond the bandwidth of the feedback loop. The modulation port bandwidth of the current driver is too low to achieve a phase-lock, as are the modulation ports of several commercial laser current drivers that were tested. Thus, the modulation current is injected onto the laser anode via a $1\,\mathrm{k\Omega}$ resistor (as depicted in Fig.(\ref{fig:feedback_schematic})). This converts the voltage modulation into a small modulation of the laser current.

\section{Results and discussion}

To investigate the performance of the phase-lock, several tools are incorporated into the setup: a spectrum analyzer is used to view the twice-downconverted beatnote at the input of the phase-frequency detector as well as the error signal generated by the phase-frequency detector; an oscilloscope is used to view the once-downconverted beatnote between the two mixers; a wavemeter performs coarse measurements of the DBR laser frequency; and the Sirah EagleEye optical spectrum analyzer is used for linewidth measurements.

\subsection{Laser frequency modulation}

\begin{figure}[t]
\centering\includegraphics[width=\textwidth]{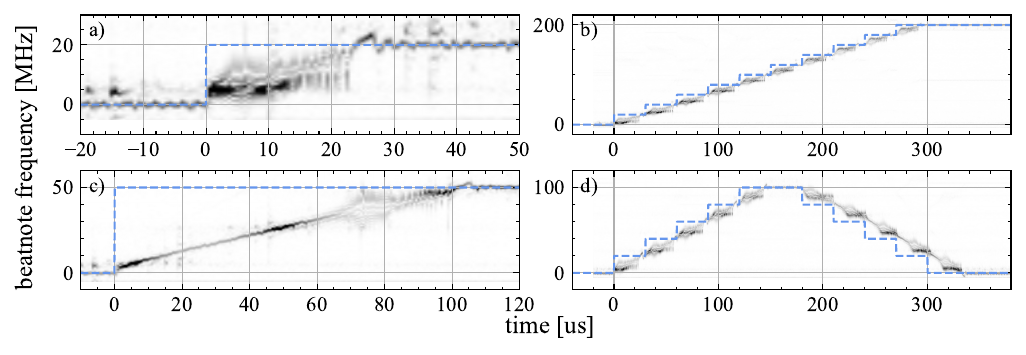}
\caption{Short-time Fourier transforms of the beatnote between the DBR laser and the ECDL reference laser for different frequency modulation schemes. The frequency axis is shifted by $-5\,\mathrm{MHz}$ so that the initial frequency is at $0\,\mathrm{MHz}$. In each panel the downconversion frequency $f_{\text{LO}_2}$ is jumped in a different way. The value of $f_{\text{LO}_2}$ as a function of time is indicated by the dashed blue line. In panels (a) and (c) $f_{\text{LO}_2}$ is jumped a single time at $t=0\,\mathrm{\mu s}$ with the standard and reduced integral gains, respectively. In panels (b) and (d) $f_{\text{LO}_2}$ is jumped several times starting at $t=0\,\mathrm{\mu s}$, up $200\,\mathrm{MHz}$ in (b) and up and then down $100\,\mathrm{MHz}$ in (d). These panels depict the versatility of the DBR laser frequency sweeps and the high rate at which the DBR frequency can be modulated. In a system with infinite feedback bandwidth the short-time Fourier transform frequency would exactly follow the dashed blue line.}
\label{fig:dbr_modulation}
\end{figure}

To investigate the frequency modulation capabilities of the phase-lock, we recorded the beatnote between the DBR laser and ECDL and increased $f_{\text{LO}_2}$ by $\Delta f$. In these tests the initial value of $f_{\text{LO}_2}$ was $255\,\mathrm{MHz}$. To retain the phase-lock during this step the beatnote frequency needed to increase by $\Delta f$. Since the beatnote frequency is several GHz, we could not record it directly. Instead we use an RF coupler to split the once-downconverted beatnote signal. Then we applied a second downconversion at $400\,\mathrm{MHz}$ to the split-off signal, obtaining a signal at $5\,\mathrm{MHz}$ before the jump and $5\,\mathrm{MHz}+\Delta f$ after the jump. This signal was recorded at $5\,\mathrm{GS/s}$ over the duration of the frequency jump to determine the behavior of the beatnote. A short-time Fourier transform was applied to the recorded time series to observe how the DBR frequency changed during the frequency jump. Note each panel in Fig.(\ref{fig:dbr_modulation}) is shifted by $-5\,\mathrm{MHz}$ to make it easier to see changes from the nominal value of $f_{\text{LO}_2}$.

In Fig.(\ref{fig:dbr_modulation}a),  $\Delta f$ was set to $20\,\mathrm{MHz}$ at $t=0\,\mathrm{\mu s}$. At the selected integral gain it took approximately $30\,\mathrm{\mu s}$ to reacquire the phase-lock. Therefore the DBR laser frequency was swept at around $0.66\,\mathrm{MHz/\mu s}$. In Fig.(\ref{fig:dbr_modulation}c) the integral gain was decreased and  $\Delta f$ was set to $50\,\mathrm{MHz}$ at $t=0\,\mathrm{\mu s}$. In this condition it took approximately $110\,\mathrm{\mu s}$ to reacquire the phase-lock, indicating a reduced sweep rate of $0.45\,\mathrm{MHz/\mu s}$.

In Fig.(\ref{fig:dbr_modulation}b), $\Delta f$ was jumped by $20\,\mathrm{MHz}$ every $30\,\mathrm{\mu s}$ for 10 cycles starting at $t=0\,\mathrm{\mu s}$, producing a linear optical frequency sweep from $0$ to $200\,\mathrm{MHz}$ in $300\,\mathrm{\mu s}$. In Fig.(\ref{fig:dbr_modulation}d) $\Delta f$ was jumped by $20\,\mathrm{MHz}$ every $30\,\mathrm{\mu s}$ for 5 cycles (again starting at $t=0\,\mathrm{\mu s}$), followed by a delay of $60\,\mathrm{\mu s}$. Then, $\Delta f$ was jumped by $-20\,\mathrm{MHz}$ every $30\,\mathrm{\mu s}$ for 5 cycles. This sequence produced a triangle-wave like optical frequency sweep. Using a direct digital synthesizer (DDS) capable of frequency sweeps instead of discrete frequency jumps would allow for smoother transitions in frequency.

This data demonstrates that a variety of frequency modulation schemes can be implemented with our phase-lock (Fig.(\ref{fig:dbr_modulation}a) and Fig.(\ref{fig:dbr_modulation}b) in particular demonstrate the instantaneous response of the lock). Further optimization of the PI controller, or adding the capability of preemptively jumping the PI output could increase the frequency modulation rate. Also, by using a ramping DDS instead of jumping the frequency of the DDS, it may be possible to sweep the frequency without losing the phase lock for a few microseconds.

\subsection{RMS phase error}
An oscilloscope is used to measure the error signal voltage, which is then converted to the phase error using the known conversion factor between the output of the phase-frequency detector and phase error ($0.16\,\mathrm{V/rad}$). The RMS phase difference is $1.4$ radians. Since $1.4$ radians is not small compared to $\pi$, one might expect that $2\pi$ cycle slips are relatively common. However, because the phase-frequency detector contains a $1/4$ frequency divider, the phase-frequency detector only sees an RMS phase difference of $0.35$ radians, and cycle slips are not observed.

\subsection{Beatnote measurement}
\begin{figure}[t]
\centering\includegraphics[width=\textwidth]{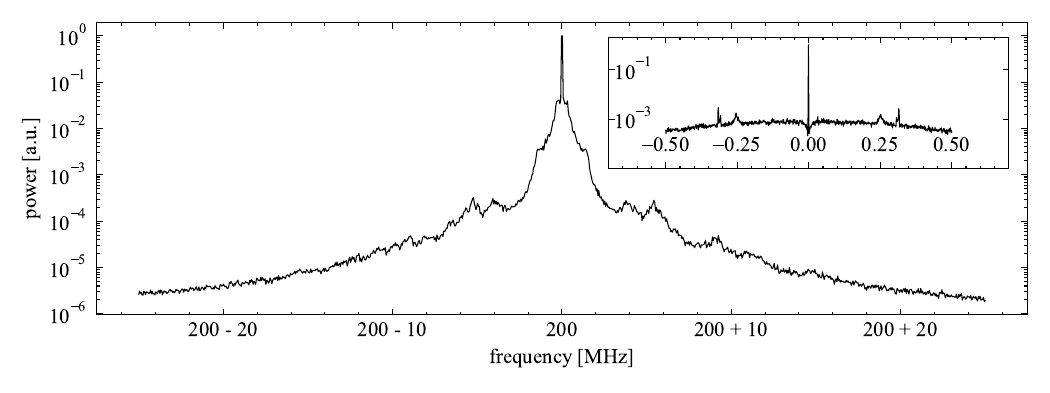}
\caption{Phase-locked downconverted beatnote measured on a spectrum analyzer with $f_\text{RF}=200\,\mathrm{MHz}$. The data was recorded with a resonant bandwidth of $10\,\mathrm{kHz}$ at 1000 points over the $50\,\mathrm{MHz}$ span. The inset is recorded with a resonant bandwidth of $300\,\mathrm{Hz}$ at 1000 points over the $1\,\mathrm{MHz}$ span. The peak width was limited by the resonant bandwidth of the spectrum analyzer in both measurements, which indicates that the peak is a delta function (a sign of a phase-lock).}
\label{fig:dbr_beatnote}
\end{figure}

A simpler downconversion scheme was used to measure the spectrum of the beatnote signal: the beatnote was downconverted once with a $8747\,\mathrm{MHz}$ source and $f_\text{RF}$ was set to $200\,\mathrm{MHz}$. A coupler between the low pass filter and the phase-frequency detector split the beatnote onto a spectrum analyzer. The recorded spectrum is pictured in Fig.(\ref{fig:dbr_beatnote}). Both the main figure and the inset show a peak at exactly $200\,\mathrm{MHz}$. The width of the peak is limited by the bandwidth of the spectrum analyzer, indicating that the peak is a delta function. This is a clear sign of a working phase-lock.

\subsection{Linewidth measurement}

Another indication of a working phase-lock is a reduction of the DBR laser linewidth. This was investigated using the Sirah EagleEye optical spectrum analyzer. The ECDL linewidth was measured to be 54(12) kHz. The natural linewidth of the DBR laser was measured to be 404(46) kHz, which is reduced to 93(21) kHz when phase-locked.

\section{Discussion}

We demonstrated an electronic phase-lock that narrows the linewidth of a DBR laser. The optical setup for this lock is simple, making it robust to environmental drifts. The lock is versatile, allowing the laser frequency to be quickly and smoothly modulated. The electronics easy to adapt to different systems. For example, if the desired offset between the reference laser and the DBR laser is only a few hundred MHz, the first downconversion step is unnecessary and a lower-bandwidth photodiode can be used. Meanwhile, if the DBR is intended to output a fixed frequency, the second downconversion step can be eliminated. The setup is also robust and operates over several weeks without the need for realignment. This robustness stems from the fact that the phase detector circuit is insensitive to changes in the beatnote amplitude. The feedback circuits function properly provided that a few milliwatts of light from each laser remains coupled into the same fiber; therefore, day-to-day drifts in the fiber-coupled power do not degrade lock performance.

\begin{backmatter}
\bmsection{Funding}
 This research was supported by the US Department of Energy under contract DE-AC02-05CH11231, the National Science Foundation award 1806583, and NASA Jet Propulsion Laboratory award 1659506.

\bmsection{Acknowledgment}
We thank James Norby, senior sales territory manager at MKS Spectra-Physics, for lending us the Sirah EagleEye for linewidth measurements.

\bmsection{Disclosures}
The authors declare no conflicts of interest.

\bmsection{Data Availability}
Data underlying the results presented in this paper are not publicly available at this time but may be obtained from the authors upon reasonable request.

\end{backmatter}

\bibliography{sample}

\begin{thebibliography}{10}
\newcommand{\enquote}[1]{``#1''}

\bibitem{Clad2009}
P.~Cladé, S.~Guellati-Khélifa, F.~Nez, and F.~Biraben, \enquote{Large momentum beam splitter using {Bloch} oscillations,} {\protect\JournalTitle{Physical Review Letters}} \textbf{102} (2009).

\bibitem{Ferrari2006}
G.~Ferrari, N.~Poli, F.~Sorrentino, and G.~M. Tino, \enquote{Long-lived {Bloch} oscillations with bosonic {Sr} atoms and application to gravity measurement at the micrometer scale,} {\protect\JournalTitle{Physical Review Letters}} \textbf{97} (2006).

\bibitem{Peik1997}
E.~Peik, M.~Ben~Dahan, I.~Bouchoule, \emph{et~al.}, \enquote{{Bloch} oscillations of atoms, adiabatic rapid passage, and monokinetic atomic beams,} {\protect\JournalTitle{Physical Review A}} \textbf{55}, 2989–3001 (1997).

\bibitem{Treutlein2001}
P.~Treutlein, K.~Y. Chung, and S.~Chu, \enquote{High-brightness atom source for atomic fountains,} {\protect\JournalTitle{Physical Review A}} \textbf{63} (2001).

\bibitem{Lett1988}
P.~D. Lett, R.~N. Watts, C.~I. Westbrook, \emph{et~al.}, \enquote{Observation of atoms laser cooled below the {Doppler} limit,} {\protect\JournalTitle{Physical Review Letters}} \textbf{61}, 169–172 (1988).

\bibitem{Ishida1991}
O.~Ishida, \enquote{Lightwave frequency tracking with a tunable {DBR} laser,} {\protect\JournalTitle{Journal of Lightwave Technology}} \textbf{9}, 1083–1093 (1991).

\bibitem{Ristic2010}
S.~Ristic, A.~Bhardwaj, M.~Rodwell, \emph{et~al.}, \enquote{An optical phase-locked loop photonic integrated circuit,} {\protect\JournalTitle{Journal of Lightwave Technology}} \textbf{28}, 526–538 (2010).

\bibitem{Friederich2010}
F.~Friederich, G.~Schuricht, A.~Deninger, \emph{et~al.}, \enquote{Phase-locking of the beat signal of two distributed-feedback diode lasers to oscillators working in the {MHz} to {THz} range,} {\protect\JournalTitle{Optics Express}} \textbf{18}, 8621 (2010).

\bibitem{reduced_linewidth_cavity}
M.~Ohtsu, \enquote{Linewidth reduction of a semiconductor laser by electrical feedback,} in \emph{Conference on Lasers and Electro-Optics,}  (Optica Publishing Group, 1985), p. ThZ5.

\bibitem{Ohtsu_1985}
M.~Ohtsu and S.~Kotajima, \enquote{Linewidth reduction of a 1.5 µm {InGaAsP} laser by electrical feedback,} {\protect\JournalTitle{Japanese Journal of Applied Physics}} \textbf{24}, L256 (1985).

\bibitem{reduced_linewidth_interferometer_1}
F.~K\'{e}f\'{e}lian, H.~Jiang, P.~Lemonde, and G.~Santarelli, \enquote{Ultralow-frequency-noise stabilization of a laser by locking to an optical fiber-delay line,} {\protect\JournalTitle{Opt. Lett.}} \textbf{34}, 914--916 (2009).

\bibitem{reduced_linewidth_interferometer_2}
H.~Jiang, F.~K\'{e}f\'{e}lian, P.~Lemonde, \emph{et~al.}, \enquote{An agile laser with ultra-low frequency noise and high sweep linearity,} {\protect\JournalTitle{Opt. Express}} \textbf{18}, 3284--3297 (2010).

\bibitem{reduced_linewidth_interferometer_3}
W.-K. Lee, C.~Y. Park, J.~Mun, and D.-H. Yu, \enquote{Linewidth reduction of a distributed-feedback diode laser using an all-fiber interferometer with short path imbalance,} {\protect\JournalTitle{Review of Scientific Instruments}} \textbf{82}, 073105 (2011).

\bibitem{reduced_linewidth_interferometer_4}
A.~Sivananthan, H.~chul Park, M.~Lu, \emph{et~al.}, \enquote{Integrated linewidth reduction of a tunable {SG-DBR} laser,} in \emph{CLEO: 2013,}  (Optica Publishing Group, 2013), p. CTu1L.2.

\bibitem{libbrecht_hall_steck_driver}
C.~M. Seck, P.~J. Martin, E.~C. Cook, \emph{et~al.}, \enquote{Noise reduction of a {Libbrecht–Hall} style current driver,} {\protect\JournalTitle{Review of Scientific Instruments}} \textbf{87}, 064703 (2016).

\end{thebibliography}

\end{document}